\documentclass[twocolumn,showpacs,preprintnumbers,amsmath,amssymb,nofootinbib]{revtex4}
\input psfig.sty

\usepackage{graphicx}
\usepackage{dcolumn}
\usepackage{bm}

\begin{document}

\def\be{\begin{equation}}
\def\ee{\end{equation}}
\def\lb{\label}


\title{Non-gaussian statistics and the relativistic nuclear equation of state}

\author{F. I. M. Pereira$^{1}$} \email{flavio@on.br}

\author{R. Silva$^{2,3}$} \email{raimundosilva@dfte.ufrn.br}

\author{J. S. Alcaniz$^{1}$} \email{alcaniz@on.br}

\affiliation{$^{1}$Observat\'orio Nacional, 20921-400 Rio de Janeiro RJ, Brasil}
\affiliation{$^{2}$Departamento de F\'{\i}sica - UFRN 59072-970, Natal - RN, Brasil}
\affiliation{$^{3}$Departamento de F\'{\i}sica - UERN 59610-210, Mossor\'o - RN, Brasil}

\date{\today}

\begin{abstract}

We investigate possible effects of quantum power-law statistical mechanics on
the relativistic nuclear equation of state in the context of the Walecka
quantum hadrodynamics theory. By considering the Kaniadakis non-Gaussian
statistics, characterized by the index $\kappa$ (Boltzmann-Gibbs entropy is
recovered in the limit $\kappa\rightarrow 0$), we show that  the scalar and
vector meson fields become more intense due to the non-Gaussian statistical
effects ($\kappa \neq 0$). From an analytical treatment, an upper bound on $\kappa$ ($\kappa < 1/4$) is found. We also show that as the parameter $\kappa$ increases the nucleon effective mass diminishes and the equation of state becomes stiffer. A  possible connection between phase transitions in nuclear matter and the $\kappa$-parameter is largely discussed.

\end{abstract}

\maketitle

\section{Introduction}\label{sec1}
In the theoretical treatment of the properties of nuclear matter, the relativistic phenomenological approach developed by Walecka~\cite{SW}, the so-called quantum hadrodynamics (QHD-I), represents one of the important approaches to the highly nonlinear behavior of strong interactions at the hadronic energy scales. This model provides a thermodynamically consistent theoretical framework for the description of bulk static properties of strong interacting many-body nuclear systems \cite{text1}. Formally, QHD-I is a strong-coupling renormalizable field theory of nucleons interacting via the exchange of (isoscalar) scalar ($\sigma$) and vector($\omega$) mesons \cite{SW}. The model has been largely used in calculations of nuclear matter and finite nuclei (see, e.g., \cite{Review} and Refs. therein). 
 
An important aspect worth emphasizing concerning the Walecka treatment is that it is based on standard quantum statistical relations, i.e., Fermi-Dirac distributions. On the other hand, as is well known, some restrictions to the applicability of the standard statistical mechanics have motivated investigations of power-law or non-gaussian statistics, both from theoretical and experimental viewpoints. In this concern, the Tsallis nonextensive statistical mechanics \cite{tsallis1} and the extensive generalized power-law statistics developed by Kaniadakis~\cite{k1} are the most investigated frameworks. Several consequences (in different branches) of the former framework have been investigated in the literature \cite{todos}, which includes systems of interest in high energy physics, namely, the problem of solar neutrino \cite{kaniadakis96}, the charm quark dynamics in a thermal quark-gluon plasma for the case of collisional equilibration \cite{walton00}, interpretations for central Au-Au collisions at RHIC energies in a Relativistic Diffusion Model (RDM) \cite{george04}, among others (see, e.g., \cite{tsallis1}). The Kaniadakis non-gaussian statistics in turn is characterized by the $\kappa$-entropy (see Sec. III) that emerges naturally in the framework  of the so-called {\it kinetic interaction principle}~\cite{k1}. Several physical features  of the $\kappa$-distribution have also been theoretically investigated as, for instance, the self-consistent relativistic
statistical theory \cite{k1}, nonlinear kinetics \cite{KaniaH01}, and the H-theorem from a generalization of the chaos molecular hypothesis \cite{rai0506}.

In this paper, by following our previous results~\cite{fimPRC2007}, we adopt the Kaniadakis non-gaussian statistics and study the effects of the $\kappa$-generalization for the Fermi-Dirac (FD) distributions on the QHD-I theory, especially what concerns its effects on the equation of state (EoS) of the nuclear matter. We also investigate the influence of these non-gaussian effects on phase transitions in the nuclear matter, as discussed in the standard context by Theis \emph{et al.}~\cite{TSP}. 

This paper is organized as follows. In Sec. II, we present the basic formalism of the mean field theory of QHD-I, which is important for the calculation of some quantities of nuclear matter. A brief review of Kaniadakis statistics is presented in Sec. III. In Sec. IV the convergence of the calculation is discussed and it is shown that the allowed values of $\kappa$ lie in the range $0 <\kappa< 0.25$. Our main results are discussed in Sec. V. In Sec. VI, we consider the connection between the $\kappa-$generalisation of the QHD-I theory and the phase transition in nuclear matter, as discussed in the paper \cite{TSP}. We summarize our main conclusions in Sec. VII.

\section{Basics of QHD-I }\label{sec2}

The Lagrangian density describing the nuclear matter reads \cite{SW}
\begin{eqnarray}  \label{DensLagran1}
{\cal L} =\bar{\psi }[(i\gamma _\mu (\partial ^\mu
-g_\omega\omega^{\mu})-(M-g_\sigma
\sigma)]\psi \nonumber \\
 + \frac 12(\partial _{\mu} \sigma \partial ^{\mu}
\sigma-m_{\sigma}^{2} \sigma^{2})-\frac {1}{4}\omega_{\mu \nu}
\omega^{\mu \nu}
 + \frac{1}{2}m_{\omega}^{2}\omega_{\mu} \omega^{\mu}~,
\end{eqnarray}
which represents nuclear matter composed by nucleons coupled to two mesons, namely, the $\sigma$ and $\omega$ mesons (for details see reference \cite{SW}).

Applying standard techniques from field theory  and the mean-field approach, we obtain the scalar density
\begin{equation}
\varrho_S=\frac{\gamma_{\rm N}}{(2\pi)^3}\int\frac{M^*}{E^*(k)}[n(\nu,T)+{\bar{n}}(\nu,T)]d^{3}k~,
\label{RS}
\end{equation}
where $M^*$ is the effective mass
\begin{equation}
M^*=M - g_{\sigma}\sigma = M-\frac{g_\sigma^2}{m_\sigma^2}\rho_S\;.
\label{ms}
\end{equation}
The baryon number density, the energy density and pressure are given, respectively, by
\begin{equation}\label{RB}
\varrho_B=\frac{\gamma_{\rm
N}}{(2\pi)^3}\int[n(\nu,T)-{\bar{n}}(\nu,T)]d^{3}k,
\end{equation}
\begin{eqnarray}\label{edens}
\varepsilon&=&\frac{1}{2}\frac{g_\omega^2}{m_\omega^2}\varrho_B^2
+\frac{1}{2}\frac{m_\sigma^2}{g_\sigma^2}(M-M^*)^2+\nonumber\\
&&\frac{\gamma_{\rm N}}{(2\pi)^3}\int{E^*(k)}
[n(\nu,T)+{\bar{n}}(\nu,T)]d^{3}k\;,
\end{eqnarray}
\begin{eqnarray}\label{press}
p&=&\frac{1}{2}\frac{g_\omega^2}{m_\omega^2}\varrho_B^2
-\frac{1}{2}\frac{m_\sigma^2}{g_\sigma^2}(M-M^*)^2+\nonumber\\
&&\frac{1}{3}\frac{\gamma_{\rm N}}{(2\pi)^3}\int\frac{k^{2}}{E^*(k)}
[n(\nu,T)+{\bar{n}}(\nu,T)]d^{3}k\;,
\end{eqnarray}
where
\begin{equation}
\label{FD}
 n(\nu,T)=\frac{1}{{\rm
e}^{\beta[E^*(k)-\nu]}+1}
\end{equation}
and
\begin{equation}
\bar{n}(\nu,T)=\frac{1}{{\rm e}^{\beta[E^*(k)+\nu]}+1},
\end{equation}
are the usual FD distributions for baryons and anti-baryons, with $E^*(k)=\sqrt{k^2+{M^*}^2}$, $\beta=1/k_BT$. The parameter
$\nu\equiv\mu-g_\omega\omega_0=\mu-(g_\omega/m_\omega)^2\varrho_B$
is the effective chemical potential, and $\gamma_N$ is the
multiplicity factor ($\gamma_N=2$ for pure neutron matter and
$\gamma_N=4$ for nuclear matter).

Additionally, to obtain the results described in Sec. V, we  use for the coupling  constants the 
values of reference \cite{SW}, namely\footnote{For the purpose of the present work, the values   
given in Eq. (\ref{cpcts}) suffices to investigate the effects of the power-law statistics in 
neutron and nuclear matter.  Variations of the coupling constants, within the acceptable
values given in current literature, do not qualitatively affect the conclusions.},
\begin{equation}
\label{cpcts}
\bigg(\frac{g_\sigma}{m_\sigma}\bigg)^2=11.798~{\rm fm^2}~{\rm~and}~
\bigg(\frac{g_\omega}{m_\omega}\bigg)^2=8.653~{\rm fm^2}~,
\end{equation}
which are fixed to give the bind energy $E_{\rm bind}=-15.75$ MeV and $k_F=1.42$ $\rm{fm}^{-1}$. 
In Sec. VI other values of $(g_\sigma/m_\sigma)^2$ are considered.

\section{Non-gaussian framework}\label{sec3}

Recent studies on the kinetic foundations of the so-called $\kappa$-statistics led to a power-law
distribution function and a $\kappa$-entropy which emerges naturally in the framework  of the {\it kinetic interaction principle} (see, e.g., Ref. \cite{k1}). Formally, the $\kappa$-framework is based on the $\kappa$-exponential and the  $\kappa$-logarithm functions which are defined as \cite{k1}
\begin{equation}\label{expk}
\exp_{\kappa}(f)= (\sqrt{1+{\kappa}^2f^2} + {\kappa}f)^{1/{\kappa}},
\end{equation}
\begin{equation}\label{expk1}
\ln_{\kappa}(f)= ({f^{\kappa}-f^{-\kappa})/2\kappa},
\end{equation}
\begin{equation}
\ln_{\kappa}(\exp_{\kappa}(f))=\exp_{\kappa}(\ln_{\kappa}(f))\equiv f.
\end{equation}
The $\kappa$-entropy associated with this $\kappa$-framework is given by
\begin{equation}\label{e1}
S_{\kappa}(f)=-\int d^{3}p f \ln_{\kappa}f,
\end{equation}
which fully recovers standard Boltzmann-Gibbs entropy, $S_{\kappa=0}(f)=-\int f \ln f d^3 p$, in the limit
$\kappa\rightarrow 0$.

\subsection{Quantum Statistics}\label{sec3a}

We recall the main aspects of the connections between the quantum statistics and Kaniadakis framework. Specifically, the main result is that, for values of the $\kappa$ index lying in the interval $[-1,1]$, a $\kappa$-generalized quantum distributions for fermions and bosons can be written as \cite{k1}
\begin{equation}\label{nq}
 n_\kappa(\mu,T)=\frac{1}{\tilde{e}_\kappa(\beta(\epsilon-\mu))\pm1},
  \end{equation}
where $\tilde{e}_\kappa$ reads
\begin{eqnarray}
\label{TPM}\tilde{e}_\kappa(x)=\left(\sqrt{1+\kappa^2 x^2} +\kappa x \right)^{1/\kappa} 
\end{eqnarray}
and $x=\beta(\epsilon-\mu)$. In particular, we observe that, $\tilde{e}_{-\kappa}=\tilde{e}_\kappa$, and that in the $\kappa\rightarrow 0$ limit, the standard FD distribution, $n(\mu,T)$, is recovered. As physically expected,  as $T\rightarrow0$, $n_\kappa(\mu,T)\rightarrow n(\mu,T)$. This amounts to saying that for studies of the interior of neutron stars (where, in nuclear scale, $T \simeq 0$) we do not expect any power-law statistic signature. On the other hand, in heavy ions collision experiments or in the interior of protoneutron stars, with typical stellar temperatures of several tens of MeV (1 MeV$=1.1065\times10^{10}$ K), power-law statistics effects can be relevant. In order to study such effects, in the next Section we combine Eqs.(\ref{ms})-(\ref{press}) with the generalized FD distributions given by Eqs. (\ref{nq}) and (\ref{TPM}).

\begin{figure*}[t]
\vspace{.2in}
\centerline{\psfig{figure=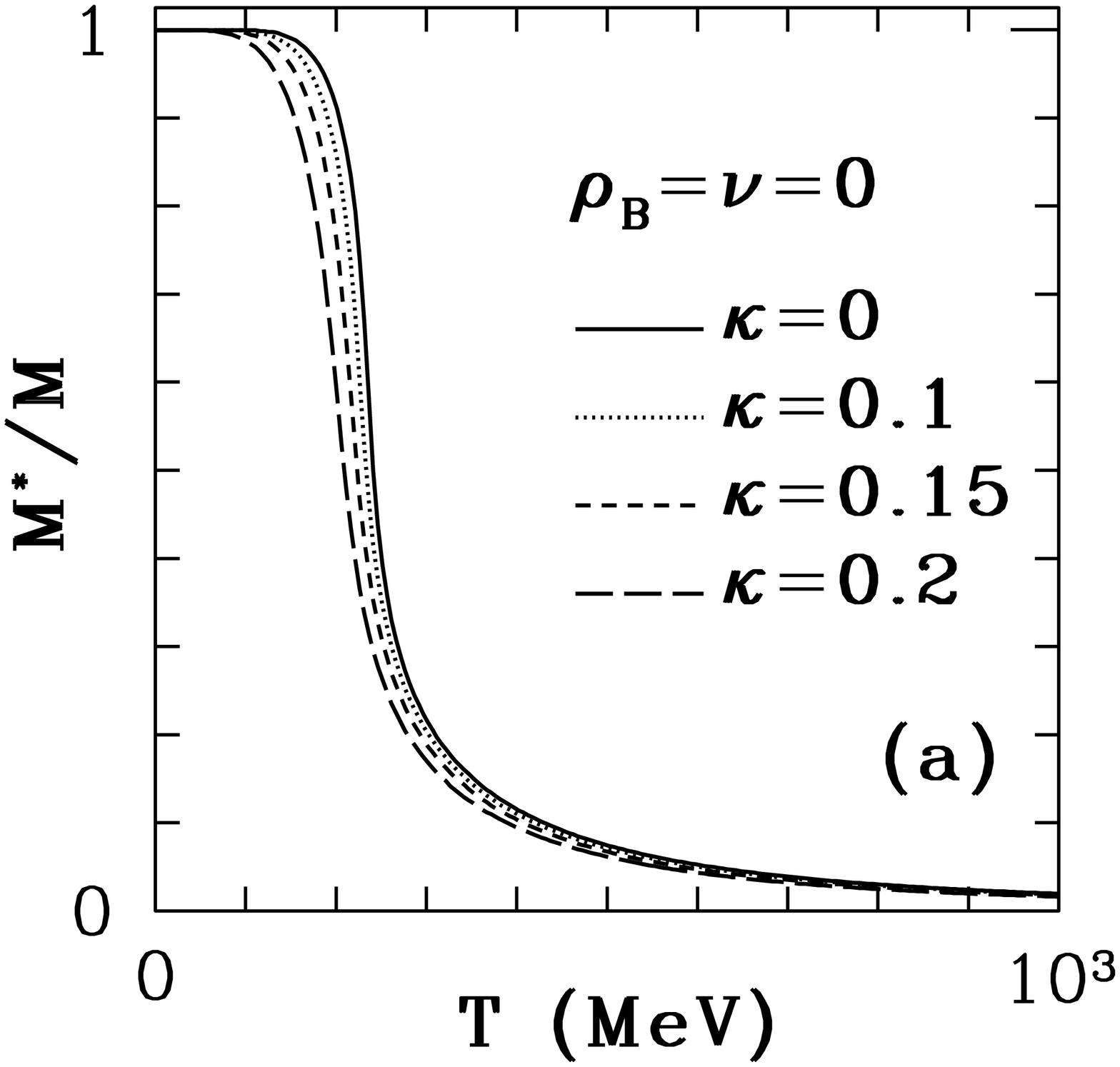,width=3.2truein,height=2.2truein}\hskip
.25in \psfig{figure=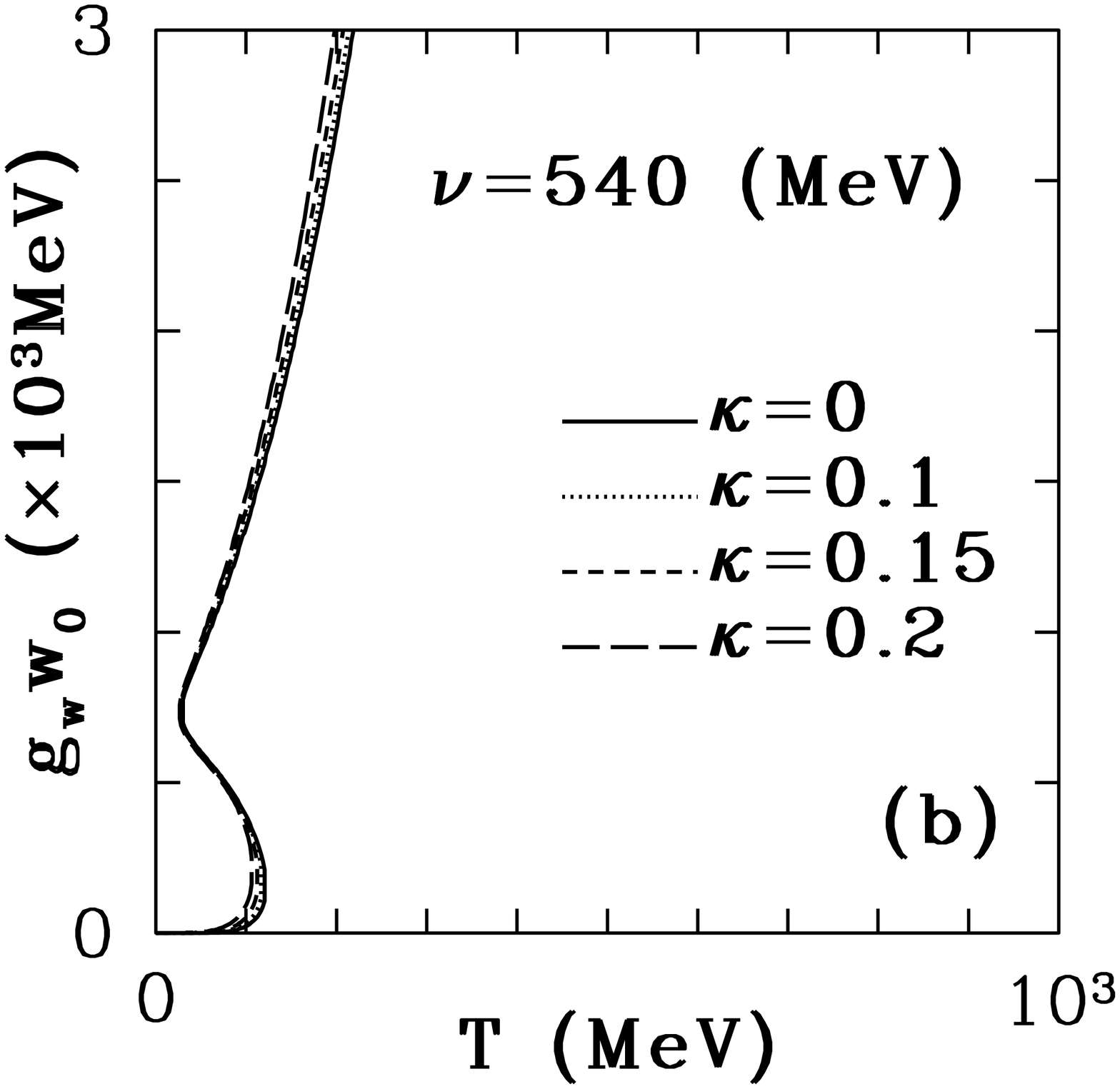,width=3.2truein,height=2.2truein}
\hskip .5in} \caption{The effective nucleon mass and the vector
meson field of pure neutron matter ($\gamma_N=2$) as function of
temperature for different values of the parameter $\kappa$. Panel (a):
the self-consistent nucleon mass at vanishing baryon density. Panel
(b): the vector meson field at nonzero baryon density corresponding
to $\nu=540$ MeV.}
\label{fig1}
\end{figure*}

\section{$\kappa$-statistics and QHD-I}\label{sec4}

Since the above function $\tilde{e}_\kappa(x)$ is a \emph{deformed} exponential function, we must verify the mathematical convergence of the integrals in Eqs. (\ref{ms})-(\ref{press}) when considering the $\kappa$-distribution (\ref{nq}). Now, taking into account that
\begin{equation}
\tilde{e}_\kappa(x)\longrightarrow (2\kappa x)^{1/\kappa},
\label{exi1}
\end{equation}
in the limit $x>>1$, we obtain the asymptotic behaviour for the integrals appearing in Eqs.(\ref{ms})-(\ref{press}), i.e.,

\begin{equation}
M^*:\int\frac{M^*~d^3k}{E^*(k)\{\tilde{e}_\kappa[E^*(k)\pm\nu]+1\}}
\longrightarrow \frac{k^2}{(2\kappa\beta)^{1/\kappa}~k^{1/\kappa}},
\label{int1}
\end{equation}

\begin{equation}
\rho_B:\int\frac{d^3k}{\tilde{e}_\kappa[E^*(k)-\nu]+1}
\longrightarrow \frac{k^3}{(2\kappa\beta)^{1/\kappa}~k^{1/\kappa}},
\label{int2}
\end{equation}

\begin{equation}
\varepsilon:\int\frac{E^*(k)~d^3k}{\tilde{e}_\kappa[E^*(k)-\nu]+1}
\longrightarrow\frac{k^4}{(2\kappa\beta)^{1/\kappa}~k^{1/\kappa}},
\label{int3}
\end{equation}

\begin{equation}
p:\int\frac{k^2~d^3k}{E^*(k)\{\tilde{e}_\kappa[E^*(k)-\nu]+1]\}}
\longrightarrow\frac{k^4}{(2\kappa\beta)^{1/\kappa}~k^{1/\kappa}},
\label{int4}
\end{equation}

\begin{figure*}[t]
\centerline{\psfig{figure=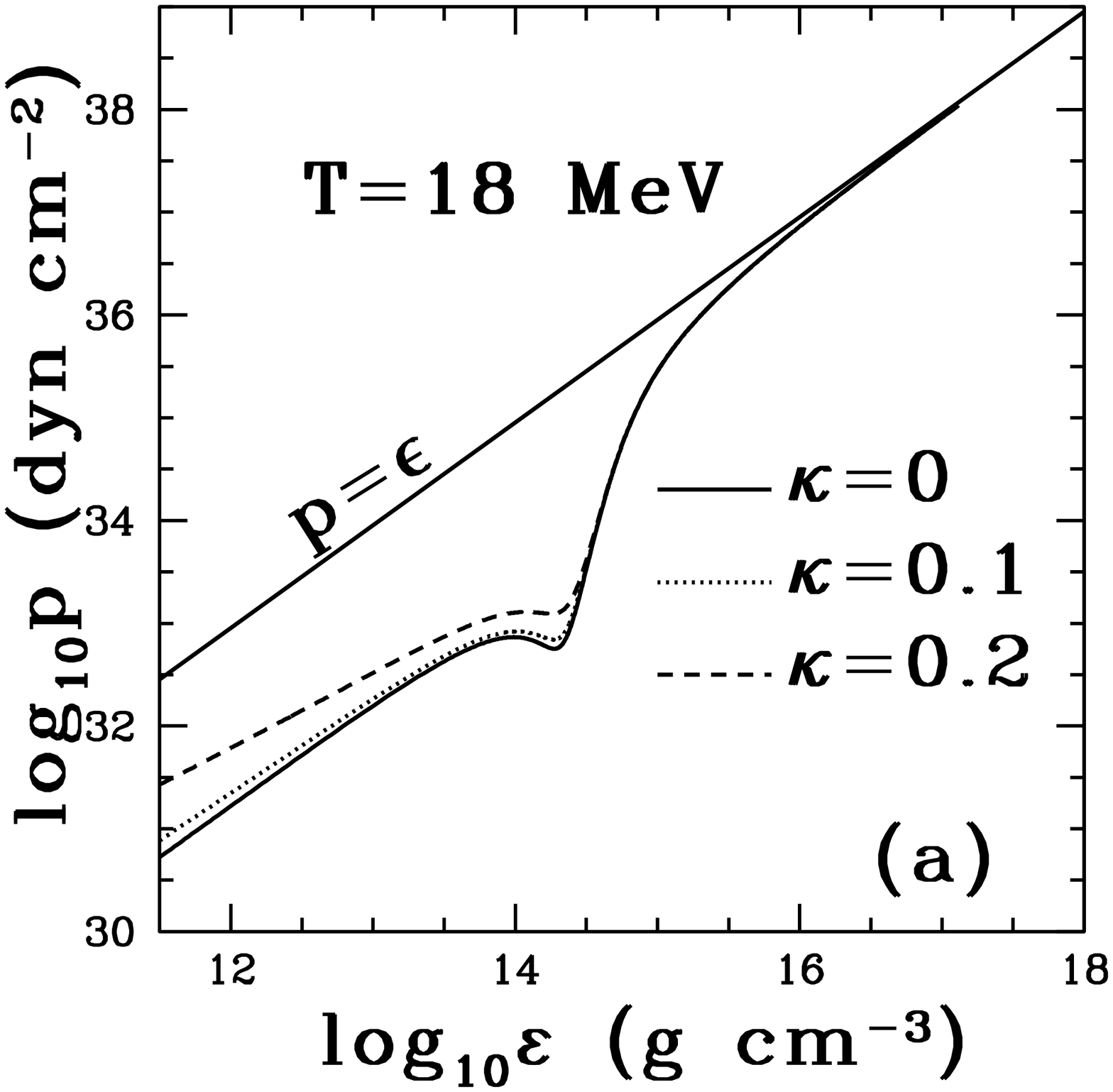,width=2.0truein,height=2.2truein}
\hspace{0.2cm}
\psfig{figure=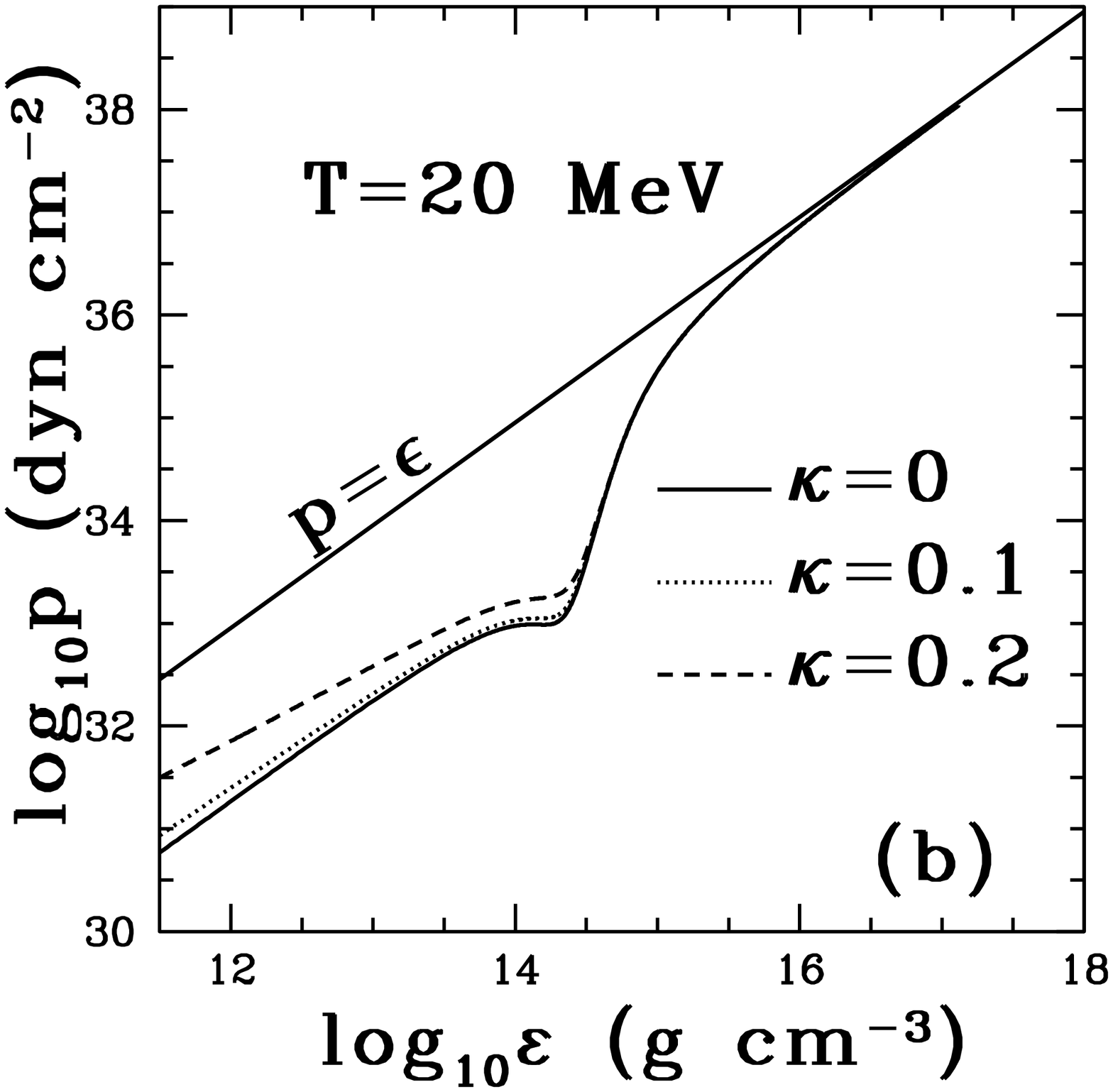,width=2.0truein,height=2.2truein}
\hspace{0.2cm}
\psfig{figure=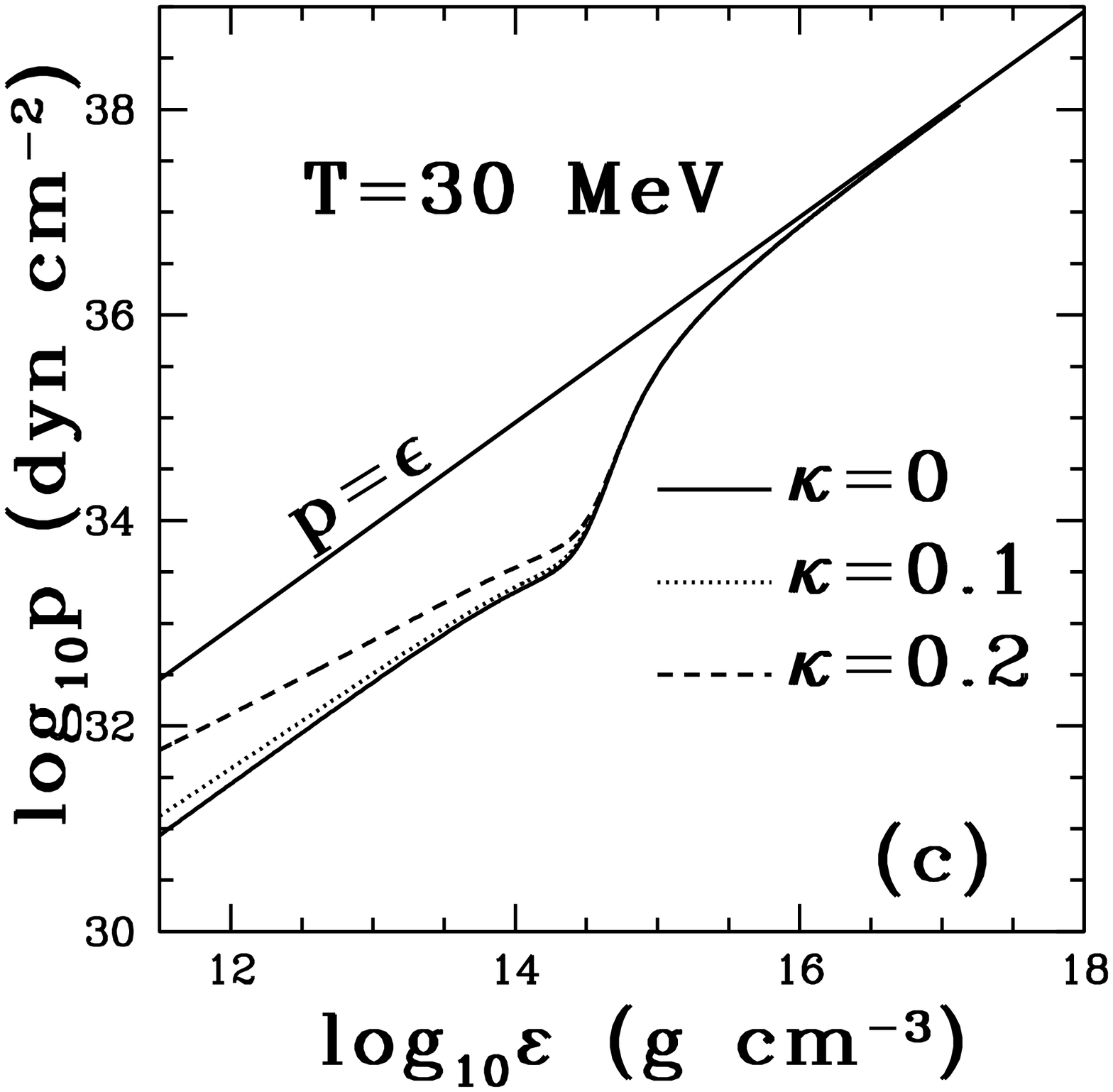,width=2.0truein,height=2.2truein}\hskip .5in}
\caption{ Isotherms of nuclear matter ($\gamma_N=4$) equation of state at  finite temperatures 
for different values of the parameter $\kappa$.}
\label{fig2}
\end{figure*}

From Eqs. (\ref{int1})-(\ref{int4}) the general asymptotic behaviour can be summarized by ${k^N}/{k^{1/\kappa}}$, for $N=2, 3, 4$.
In order to have ${k^N}/{k^{1/\kappa}}\longrightarrow0$, when
$k\longrightarrow\infty$, we find that ${1}/{\kappa}>N$, from which we obtain
\begin{equation}
~~\kappa<\frac{1}{N}~. 
\label{qM1}
\end{equation}
Note that, to simultaneously satisfy the convergence of all integrals in
Eqs.(\ref{ms})-(\ref{press}), we find that $0<\kappa<1/4$.

At high temperatures ($T\rightarrow\infty$), the analytic solution to Eq. (\ref{ms}) can be written as 
\be
M^*\rightarrow M\bigg[1+\frac{g_\sigma^2}{m_\sigma^2}
\bigg(\frac{\gamma_N}{\pi^2}\bigg)\xi_1(\kappa)(k_BT)^2\bigg]^{-1}
\lb{Mss}
\ee
where $\xi_1(\kappa)$ is given by Eq. (\ref{xin}).

Several limiting cases of the EoS are of interest:

\begin{enumerate}

\item The baryon distribution becomes a step function
$n_\kappa(\nu,0)=\theta(k_F-|{\bf k}|)$ in the limit $T\rightarrow0$ for any value of $\rho_B$.

\item The system becomes degenerate in the limit $\rho_B\rightarrow\infty$ at any $T$.

\item As $T\rightarrow\infty$ for any vaue of $\rho_B$, an equation of state similar to that of a black
body is obtained:
\be
\varepsilon\rightarrow\frac{\gamma_N}{\pi^2}\xi_3(\kappa)(k_B T)^4~,~~~p=\varepsilon/3~,
\lb{Ek}
\ee
where, in Eqs.(\ref{Mss}) and (\ref{Ek}),
\be
\xi_n(\kappa)\equiv\int_0^\infty\frac{z^n~dz}{\tilde{e}_\kappa(z)+1}~.
\lb{xin}
\ee
When $\kappa\rightarrow 0$ we have that $\xi_1(\kappa\rightarrow 0)\rightarrow\pi^2/12$
and $\xi_3(\kappa\rightarrow 0)\rightarrow7\pi^4/120$, recovering the limits of Ref. \cite{SW}.
\end{enumerate}

\begin{figure*}[tbh]
\centerline{\psfig{figure=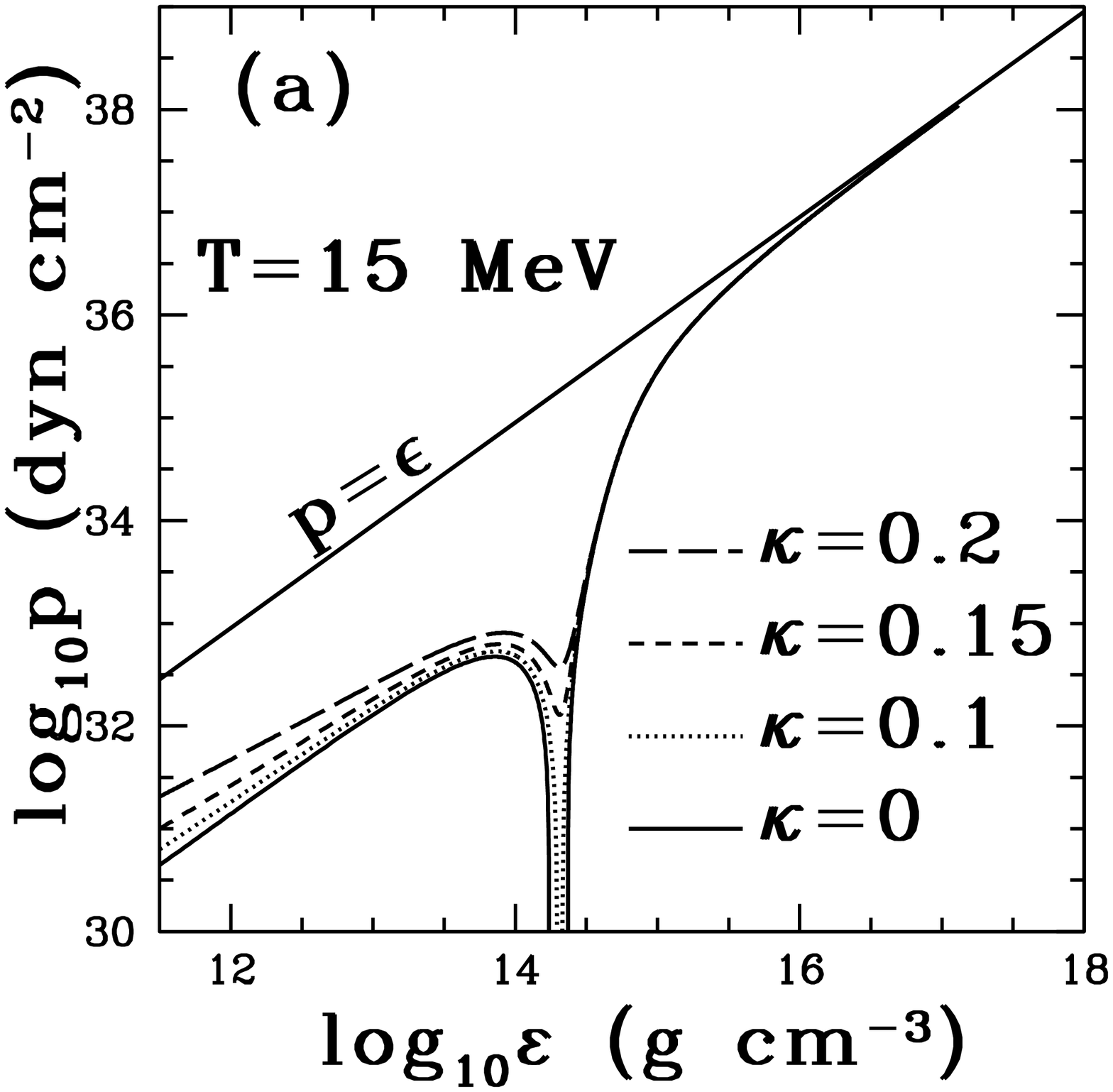,width=2.0truein,height=2.2truein}
\hspace{0.2cm}
\psfig{figure=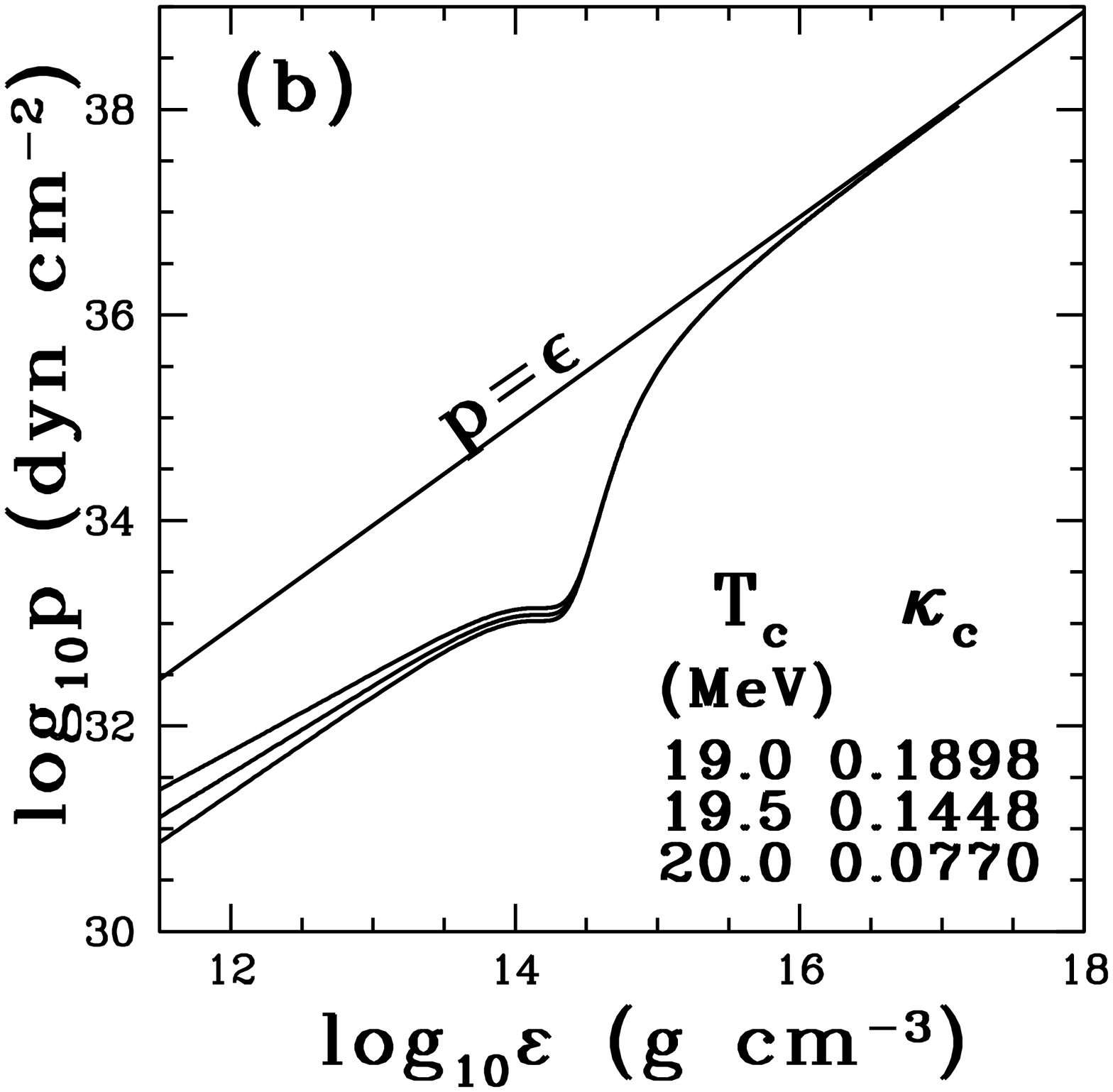,width=2.0truein,height=2.2truein}
\hspace{0.2cm}
\psfig{figure=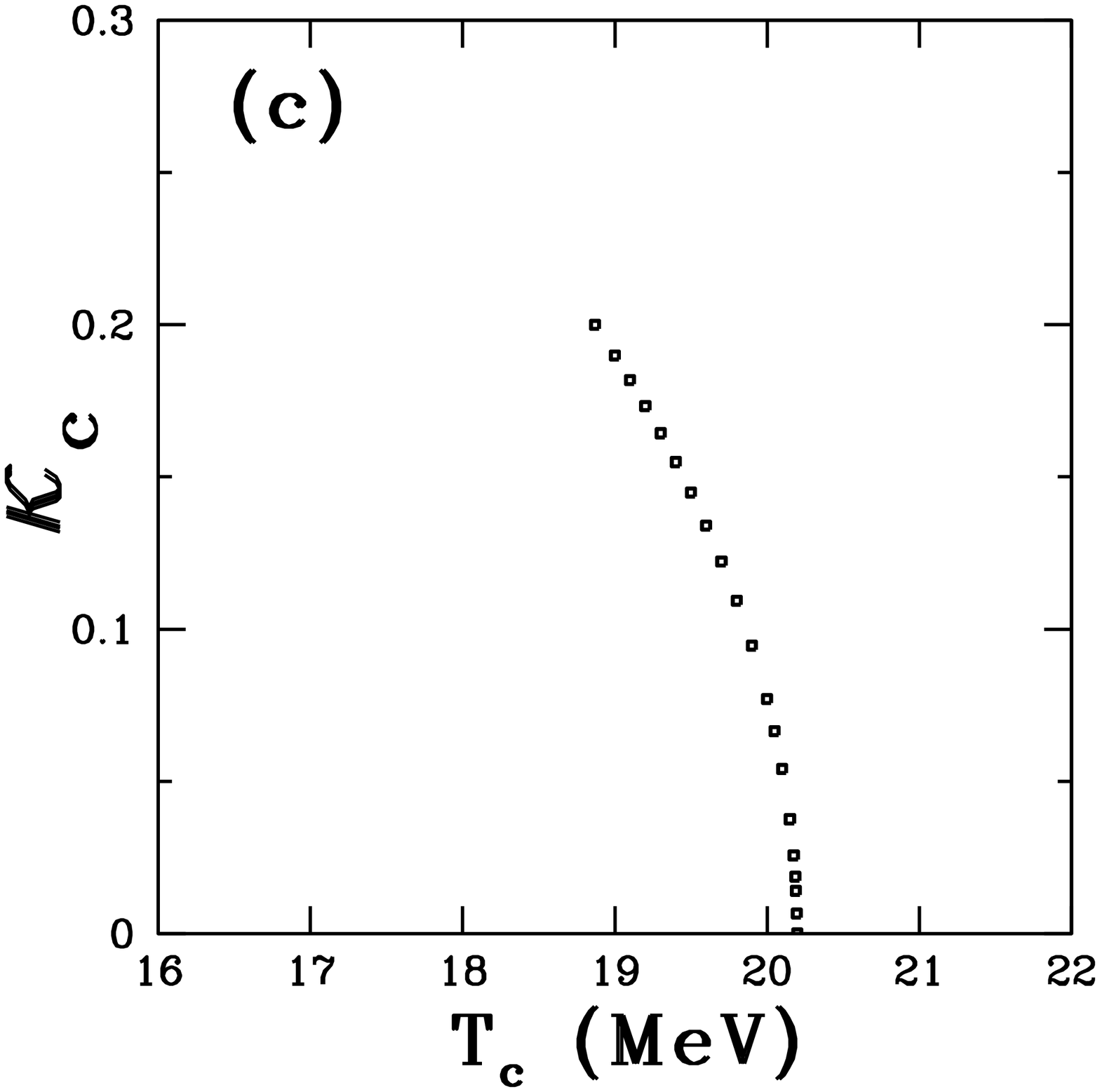,width=2.0truein,height=2.2truein}\hskip .5in}
\caption{Panel (a): Isotherms of nuclear matter ($\gamma_N=4$) at
$T=15$ MeV for several values of the parameter $\kappa$. Panel (b):
Isotherms of nuclear matter corresponding to the critical
temperatures and parameters $\kappa_c$ given in the data. The curves and
the data are in the same order from top to bottom. Panel (c): the
critical parameter $\kappa_c$ as funtion of the critical temperature in the range
$18.8\;{\rm MeV}<T_c<20.2\;{\rm MeV}$ (including the data of Panel (b)).} 
\label{fig3}
\end{figure*}

\begin{figure*}[t]
\centerline{\psfig{figure=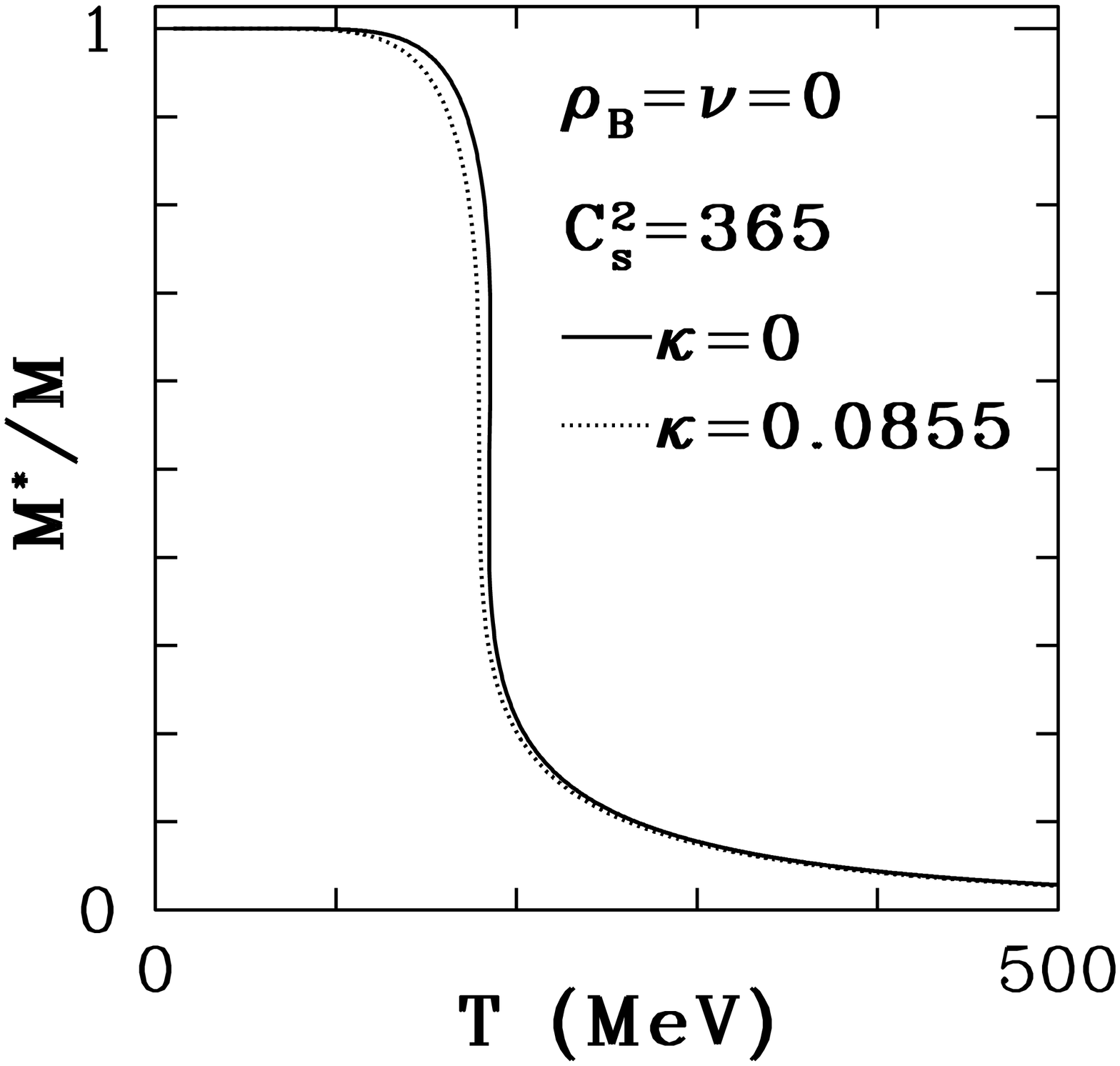,width=3.2truein,height=2.2truein}\hskip
.25in \psfig{figure=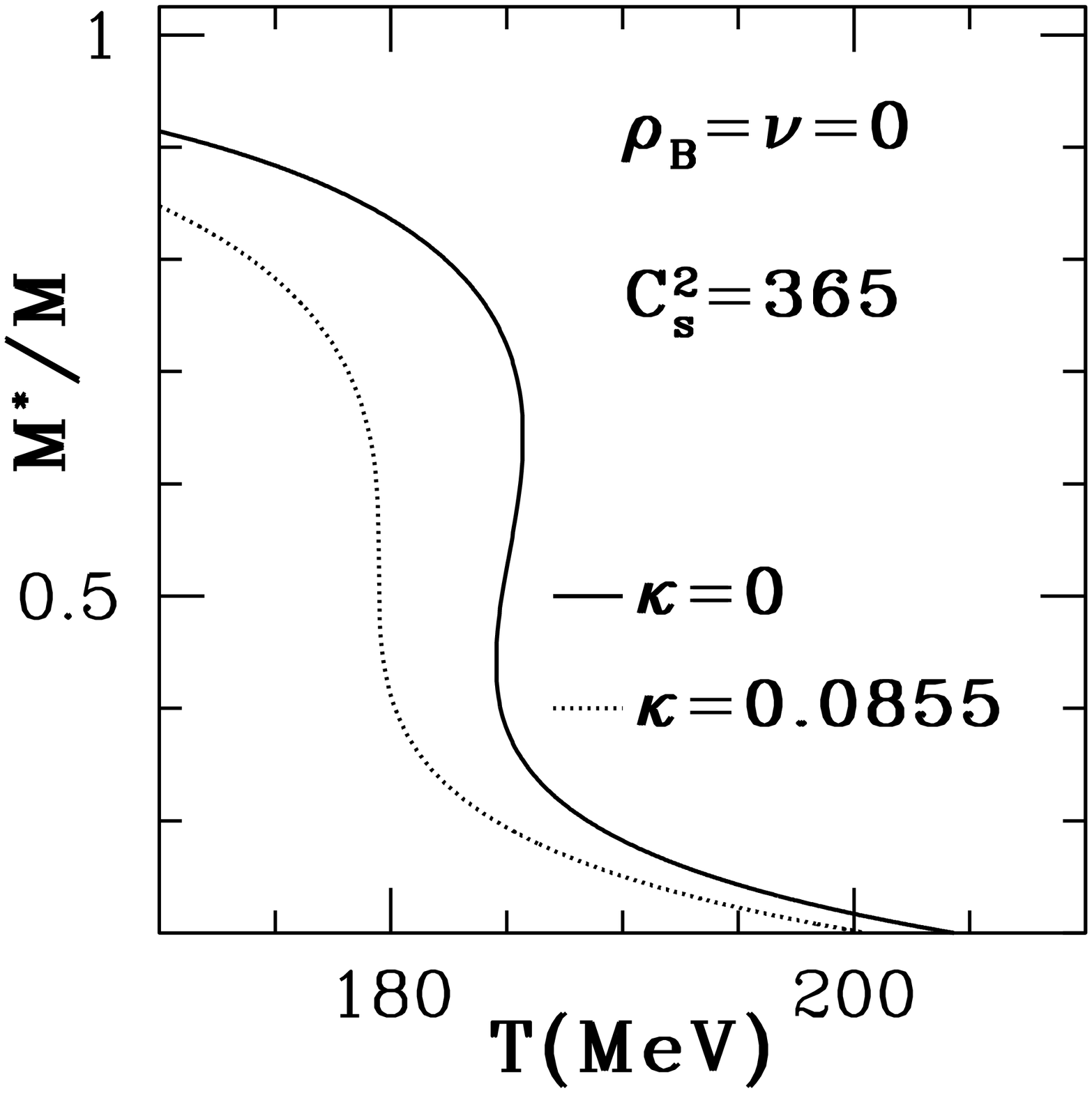,width=3.2truein,height=2.2truein}
\hskip .5in} \caption{The solution $M^*/M$ of equation of (\ref{ms}) for nuclear matter ($\gamma=4)$ 
at vanishing baryon density as a function of temperature for the same value of
$C_S^2$ and two different values of the parameter $\kappa$. The right Panel shows the same results but 
with a stretched temperature region around the point of phase transition.}
\label{fig4}
\end{figure*}

\section{the phase structure at nonzero baryon density}\label{sec5}

\subsection{EoS}

In order to study the effects of the non-gaussian framework on QHD-I theory, we have calculated numerically, for several values of temperatures and of the parameter $\kappa$, the effective nucleon mass, the vector and scalar
mesons fields for pure neutron matter ($\gamma_N=2$), as well as the EoS for nuclear matter ($\gamma_N=4$).

Fig. \ref{fig1} shows the non-gaussian effects on the effective mass and vector meson for a pure neutron matter. In Panel (a), for $\varrho_B= \nu =0$, we note that the higher the parameter $\kappa$,  the smaller the effective mass $M^*$ (and, consequently, the higher the scalar field $\sigma$, since $g_\sigma\sigma=M - M^*$). Such an effect may be physically understood in that at a given temperature, the scalar density, as source for the scalar mesons, increases with the increasing of the parameter $\kappa$. Thus, the attraction of the nucleons, mediated by the scalar mesons,  becomes stronger, reducing the effective mass. The behaviour of vector meson field $g_\omega\omega_0=(g_\omega/m_\omega)^2\varrho_B$ is shown in Panel (b).

The non-gaussian effects on the EoS of symmetric nuclear matter are shown in Fig. \ref{fig2}. The panels display the $\log{p}-\log{\epsilon}$ plane for selected values of $\kappa$. The results are ploted for arbitrarily chosen values of temperature, $T=$18 MeV, 20 MeV, and 30 MeV.  In reality, the motivation for this choice is of astrophysical interest, e.g., in the study of protoneutron stars. Clearly, the non-gaussian effect is manifested in the increasing of the pressure with the values of $\kappa$ making the EoS stiffer.

\subsection{Phase transitions}

Another effect of the power-law statistics on nuclear EoS of QHD-I concerns the phase transitions. From Panels (a) and (b) of Fig. 3, we see that the first order phase transition may be eliminated by the variation of the parameter $\kappa$. This fact can be easily visualized in Fig. \ref{fig3}{(a)}, where isotherms at $T=15$ MeV are plotted for several values of $\kappa\in[0,\;1/4)$. Note that, for increasing values of $\kappa$, the dip in the region of thermodynamical instability becomes smaller, vanishing at the turning point that defines the critical values of thermodynamical quantities ($T_c$, $p_c$, etc.). 
 
We also note that for our choice $T=15$ MeV the upper value of $\kappa$ (near the 1/4 limit discussed earlier) 
is not sufficient to eliminate the first order phase transition \cite{text}. On the other hand, first order phase 
transitions can be eliminated for values of $T$ in the region $18\;{\rm MeV}<T<20.2\;{\rm MeV}$. In this interval, 
all temperatures can be made critical. This amounts to saying that a (critical) parameter $\kappa_c$ can be determined 
in order to yield a turning point in the isotherm at a given temperature. This
is illustrated in Fig. \ref{fig3}(b) where several isotherms are displayed. InFig. \ref{fig3}(c)  
values of $\kappa_c$ are plotted as function of temperature in the range ($18.8~{\rm MeV}<T<20.2~{\rm MeV}$). 

\begin{figure*}[t]
\centerline{\psfig{figure=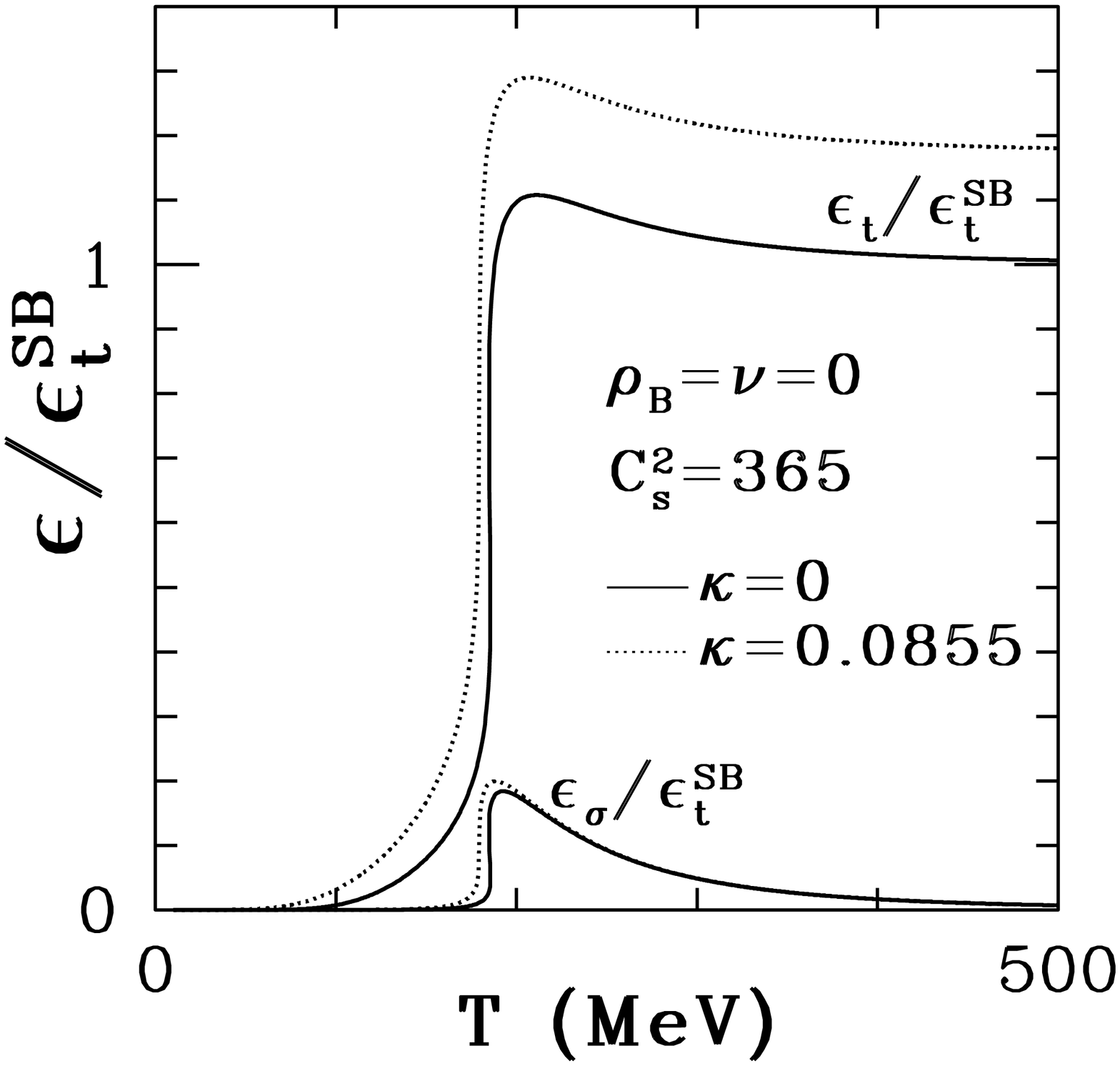,width=3.2truein,height=2.2truein}\hskip
.25in \psfig{figure=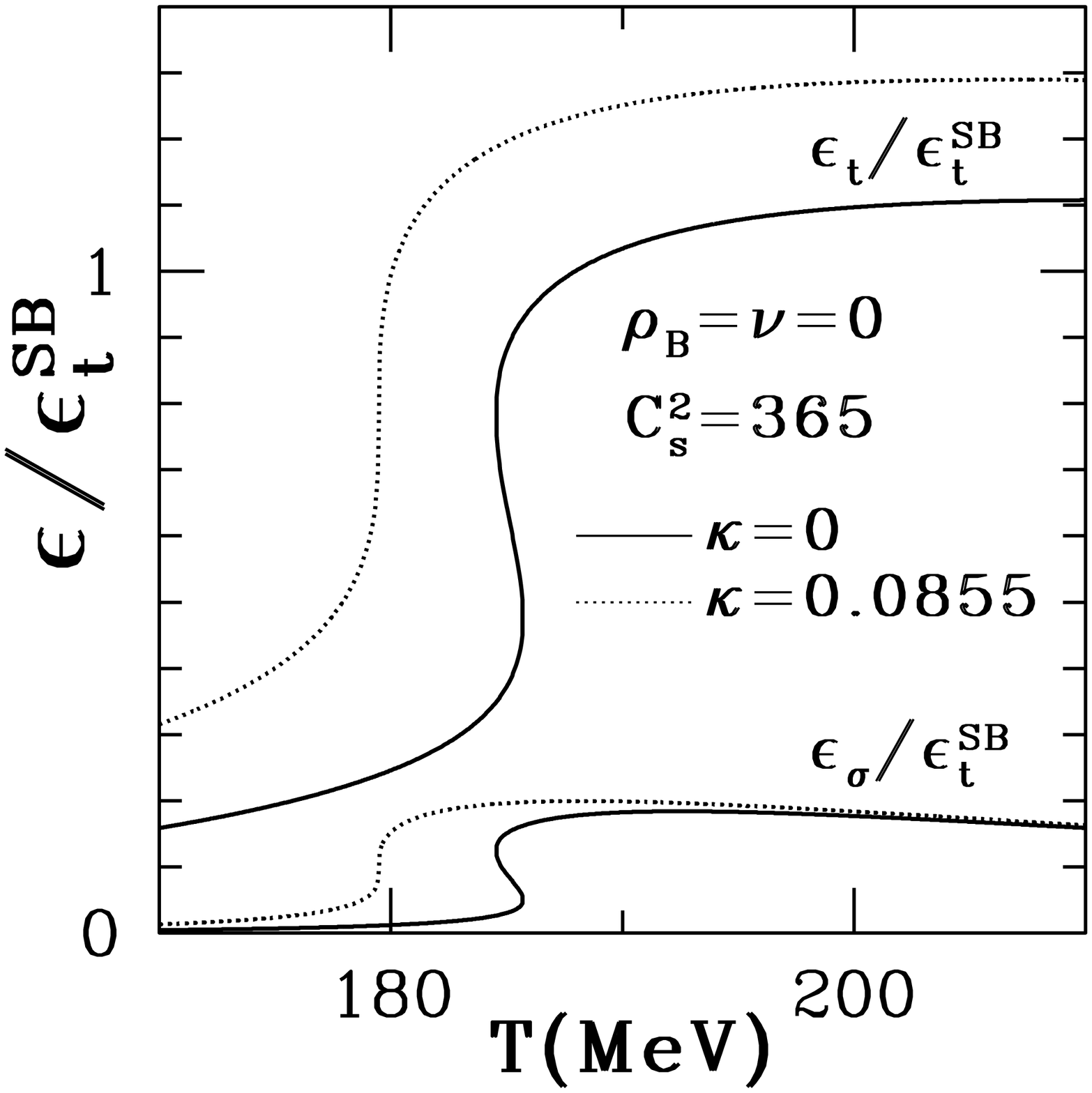,width=3.2truein,height=2.2truein}
\hskip .5in} \caption{The total energy density $\epsilon_t$ and the scalar-field energy density 
$\epsilon_\sigma$ divided by the $\kappa=0$ Stefan-Boltzmann limit $\epsilon_t^{SB}$ as a function 
of temperature, at zero baryon density of nuclear matter ($\gamma=4$). The same value of $C_S^2$ is 
considered for two different values of the parameter $\kappa$. In the right Panel the same 
results in the stretched temperature region near the transition point.}
\label{fig5}
\end{figure*}

\begin{figure*}[t]
\centerline{\psfig{figure=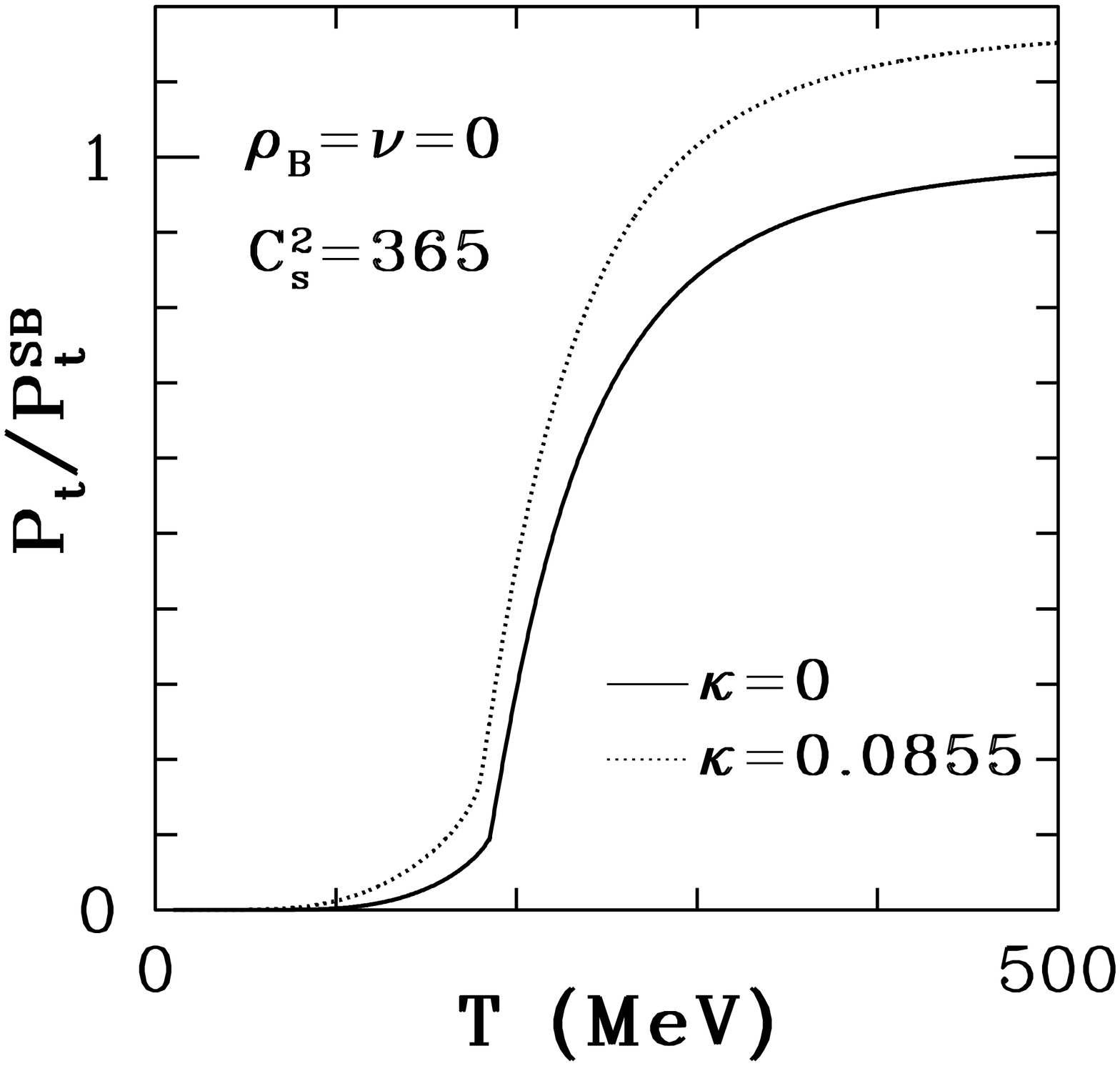,width=3.2truein,height=2.2truein}\hskip
.25in \psfig{figure=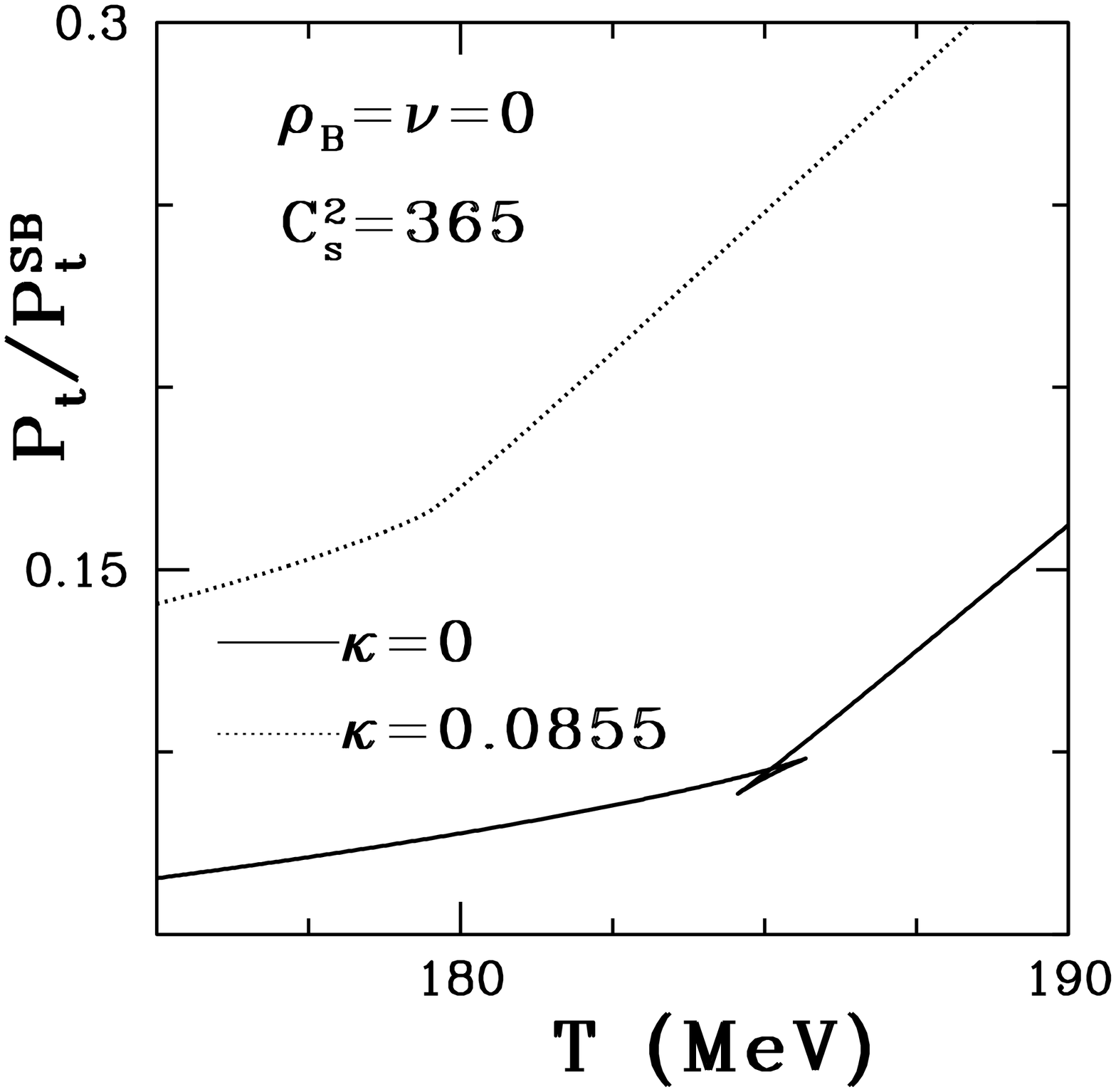,width=3.2truein,height=2.2truein}
\hskip .5in} \caption{For nuclear matter ($\gamma=4$) at vanishing baryon density, the total pressure 
divided by the corresponding $\kappa=0$ Stefan-Bolztmann limit $P_t^{SB}$ as function of temperature. 
The same value of $C_S^2$ is considered for two different values of the parameter $\kappa$. The right  
panel shows the first order phase transition point for $\kappa=0$ (solid curve). For $\kappa=0.0855$ 
(dotted curve) the phase transition is of second order.} 
\label{fig6}
\end{figure*}

\section{the phase structure at zero baryon density}\label{sec6}

Ref.~\cite{TSP} discusses experimentally a reproduction of the observed binding energy and density of nuclear matter in an area characterized by a line in the coupling-constant plane, where part of this line defines a system exhibiting a phase transition around $T_c\sim 200$ MeV. As matter of fact, a different sets of coupling constants in the coupling-constant plane were considered~\footnote{For a different sets of coupling constant, the mean field solutions provide the nuclear binding energy $-16<\varepsilon<-15 $ MeV at equilibrium densities $0.14<\rho_{\rm eq}<0.19\;{\rm fm}^{-3}$.}. 

In this Section, by considering the same arguments of Ref.~\cite{TSP}, we explore the non-gaussian phase structure of the effective Lagrangian at vanishing chemical potential and baryon density ($\varrho_B=\nu=0$). To this end, we first 
consider the range of values for the coupling constant $C_S^2$ given by 
\begin{equation}\label{Cs2}
C_S^2=(\frac{g_\sigma}{m_\sigma})^2 M^2\;
\end{equation}
in the coupling-constant plane shown in Fig.(1) of  Ref.~\cite{TSP}. At vanishing chemical potential the terms with the baryon density do not appear in Eqs. (\ref{edens}) and (\ref{press}). In what follows, we have taken as an example $C_S^2=365$.

Theis et al.~\cite{TSP} showed that the order of phase transition is strongly dependent on the actual value of the coupling constant $C_S^2$. However, in Sec.\ref{sec5}, it is shown that we can avoid Maxwell construction by the variation of the parameter $\kappa$.  Since the order of transition depends on $C_S^2$ and $\kappa$, the natural question is whether there exists some relation between $C_S^2$ and $\kappa$ at zero baryon density.

\begin{figure*}[t]
\centerline{\psfig{figure=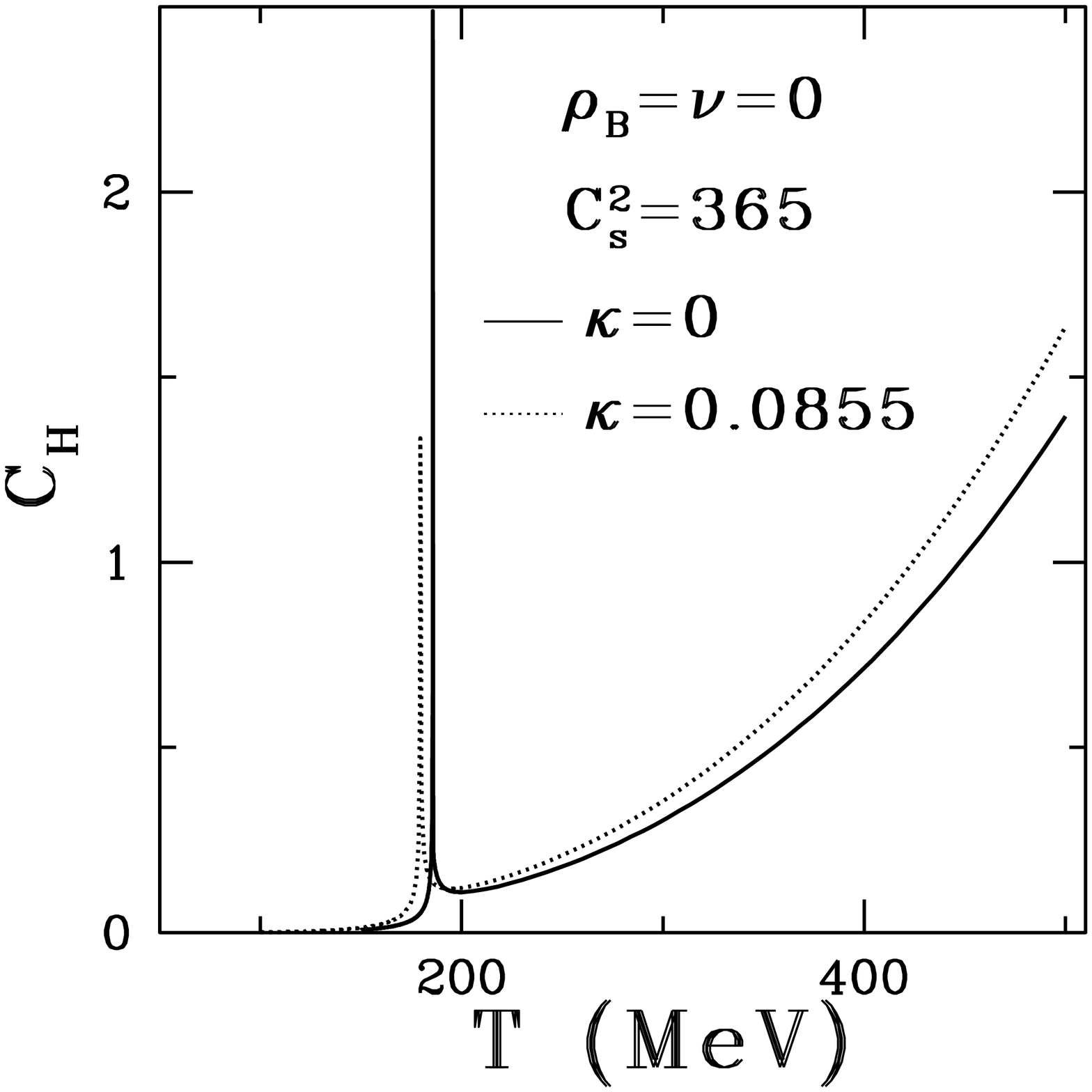,width=3.2truein,height=2.2truein}\hskip
.25in \psfig{figure=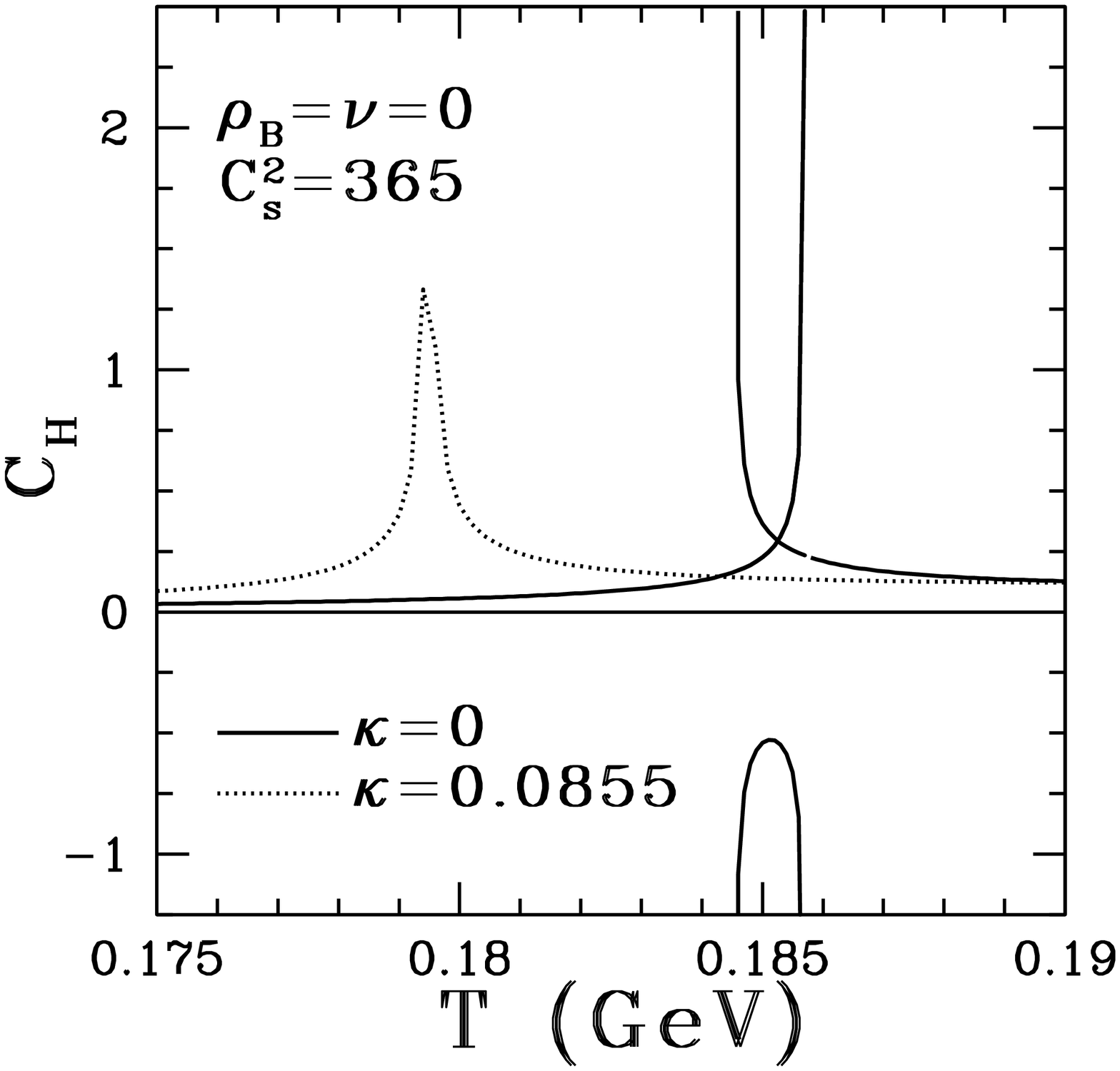,width=3.2truein,height=2.2truein}
\hskip .5in} \caption{The specific heat of nuclear matter ($\gamma=4$) at zero
baryon density divided by the corresponding $\kappa=0$ Stefan-Boltzmann limit
as function of temperature for the same value of $C_S^2$ and two different
values of the parameter $\kappa$. In the right Panel, the same results in
the stretched region around the phase transition point. For $\kappa=0$ and
$\kappa=0.0855$ the phase transitions are, respectively, of the first and second order.}
\label{fig7}
\end{figure*}

In Fig.~\ref{fig4}, the sudden drop in $M^*$ around $T\sim185$ MeV  determines the abrupt rise of the energy density and pressure.  We note that for $\kappa=0$ (Fermi-Dirac statistics) the self-consistency equation (\ref{ms}) has three solutions around $T\sim185$ MeV imposing a sudden  rise in the energy density and a peak in the specific heat. This behavior is shown in  Figs.~(\ref{fig5})-(\ref{fig7}), where the temperature dependence with total energy density, pressure and  specific heat divided by the corresponding high Stephan-Boltzmann temperature limit (for $\kappa=0$), 
given by Eq.~(\ref{Ek}), are displayed. The behavior of the curves for $\kappa=0$ characterizes a first order phase transition with the pressure curve crossing itself twice at $T\sim185$ MeV. We observe that in this region the energy density and presure are also triple valued and that the specific heat is negative. The value of $\kappa=0.0855$ was obtained by requiring that energy density and pressure to be single valued characterizing a second order phase transition with non-negative specific heat.  This allows to obtain a relation between $C_S^2$ and $\kappa$, as explained below.

\subsection{$C_S^2 - {\kappa}$ relation}

\begin{figure*}[tbh]
\centerline{\psfig{figure=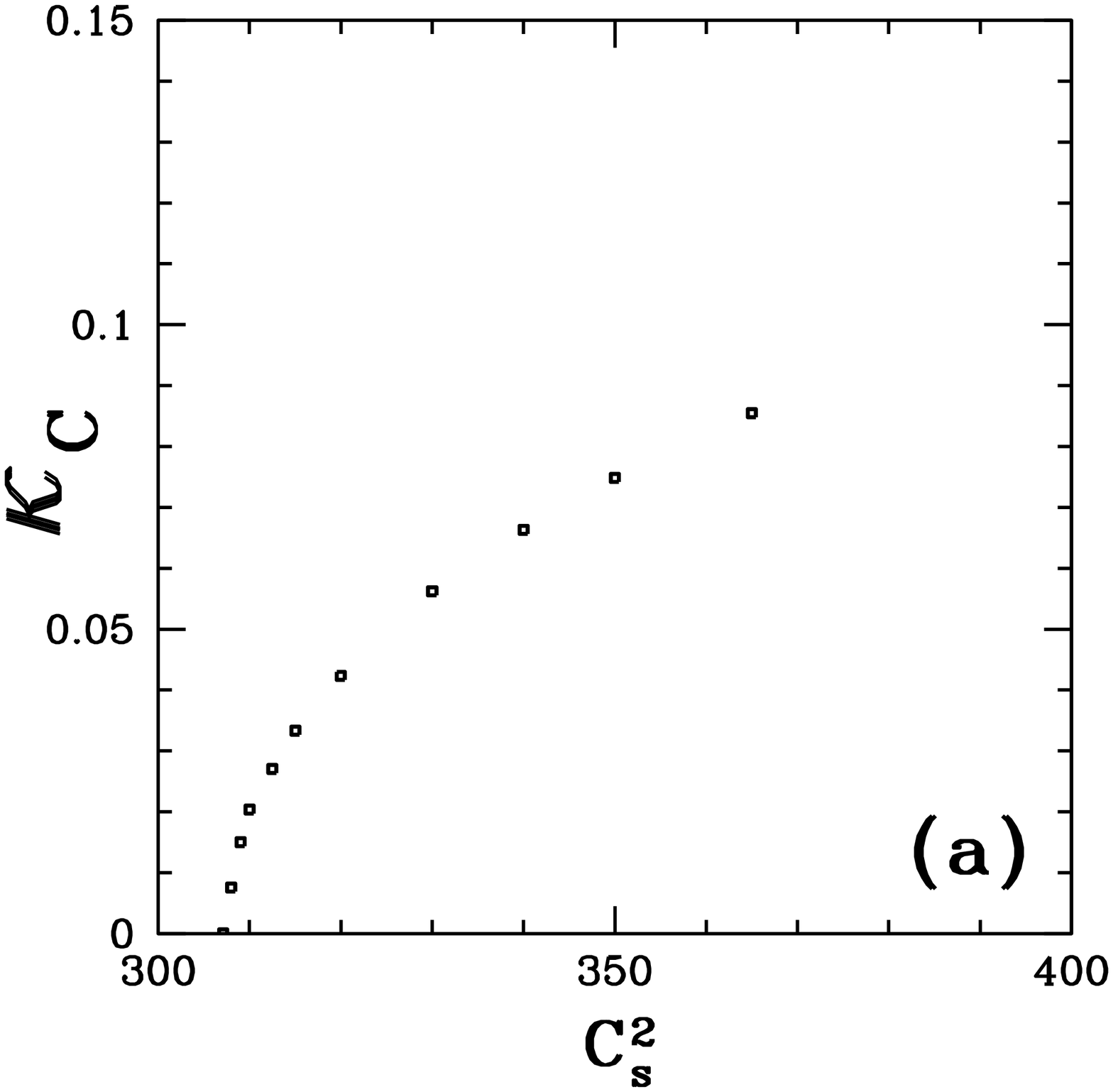,width=2.0truein,height=2.2truein}
\hspace{0.2cm}
\psfig{figure=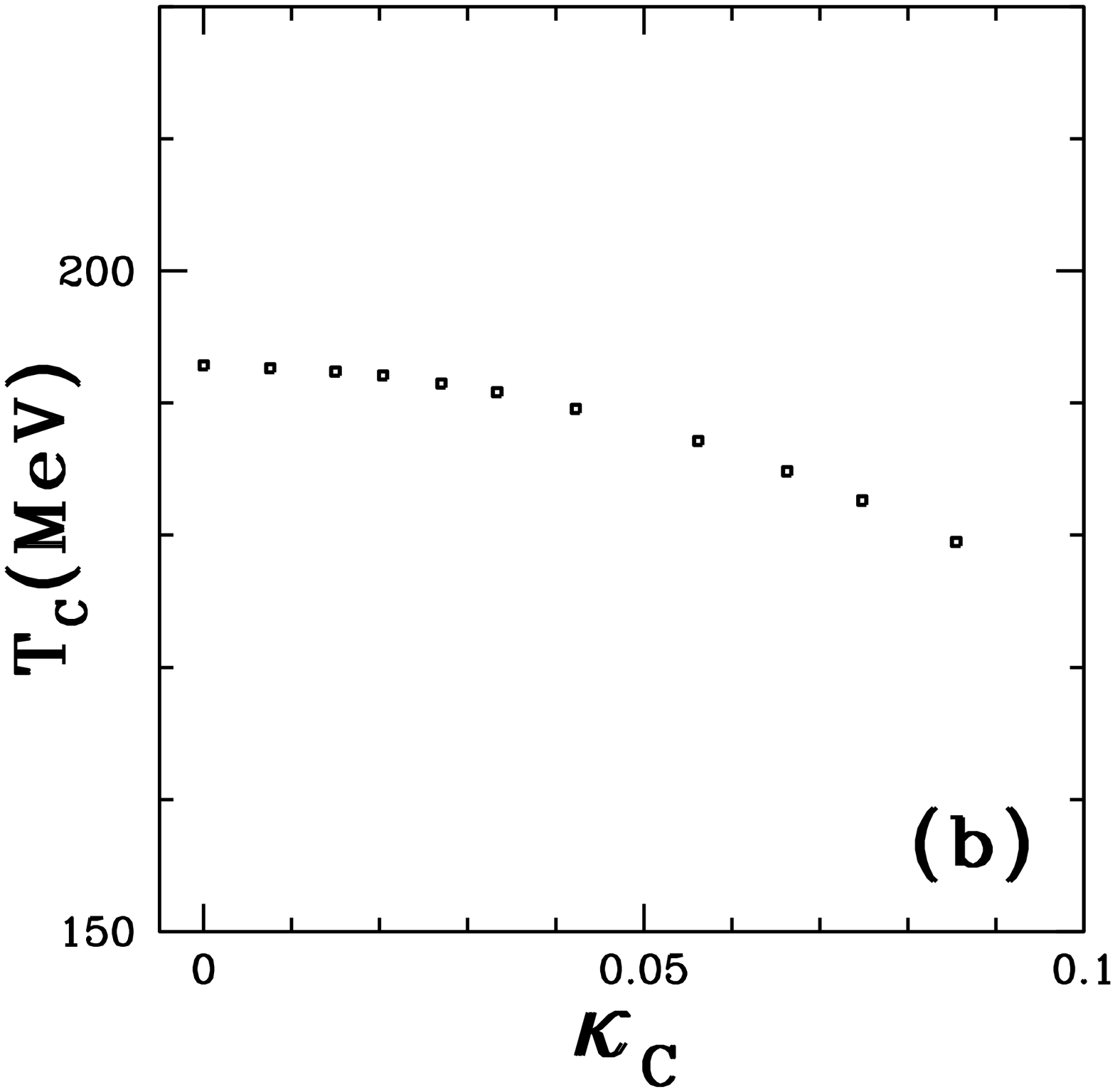,width=2.0truein,height=2.2truein}
\hspace{0.2cm}
\psfig{figure=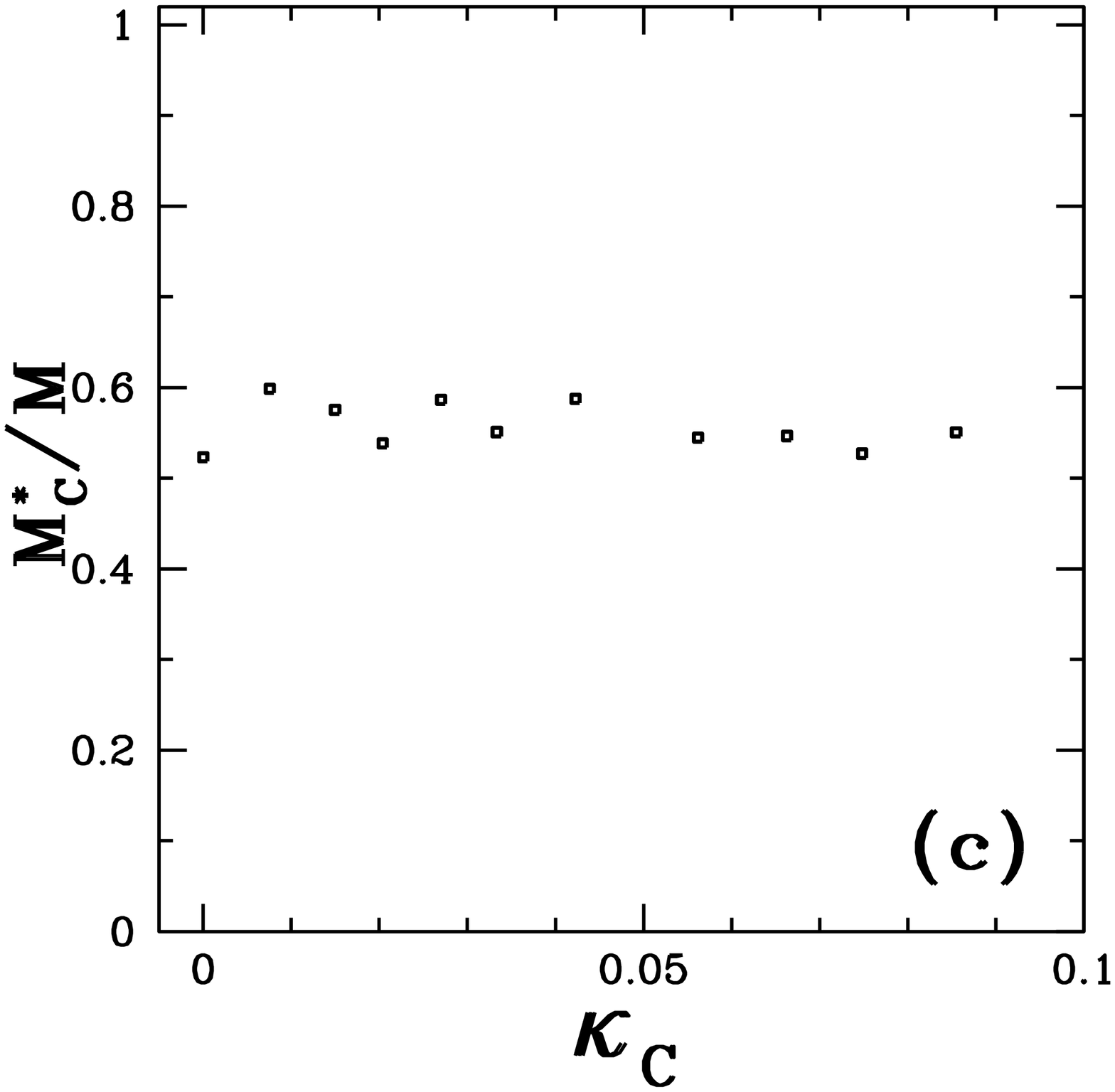,width=2.0truein,height=2.2truein}\hskip .5in}
\caption{Panel (a): Values of the critical parameter $\kappa$ for which the
phase transitions are of second order as function of the coupling constant
$C_S^2$, for nuclear matter ($\gamma=4$) at zero baryon density. 
Panel (b): The temperature corresponding to Panel (a) as function of $\kappa_C$.
Panel (c): The same as in Panel (b) but for the effective mass $M^*_C$.} 
\label{fig8}
\end{figure*}

In order to obtain such a relation, we investigate the thermodynamical behavior of the nuclear matter for several values of the coupling constant $C_S^2$ in the coupling-constant plane as follows. For each value of  $C_S^2$, the corresponding parameter $\kappa$, for which the transition is of second order, is determined. Thus, a curve in the $\kappa \times C_S^2$ plane is obtained as shown in Fig.~\ref{fig8}(a). Below this curve the phase transitions are of the first order and above it the thermodynamical behavior is smooth. Let us now show how the calculation is made via specific heat.

Differently from the treatment discussed in  Ref.~\cite{TSP}, the mathematical structure of the self-consistency
equation in our approach is not simple, so that the calculation must be done numerically. We observe that the specific heat calculated from Eq.~(\ref{edens}) is linear in $dM^*/dT$. So, whenever there is a sudden fall in $M^*$, we see a peak in the specific heat. By writing
\begin{equation}\label{Ch}
C_H=\frac{d\varepsilon}{dT}=\frac{d\varepsilon}{dM^*}\frac{dM^*}{dT}
\end{equation}
we can see from Eq. (\ref{ms}) that 
\begin{equation}\label{dMdT1}
\frac{dM*}{dT}=\frac{-2C_{M^*}M^{*3}\int_0^\infty\frac{2k^2+M^{*2}}{E^*(k)}\; n^{\{\kappa\}}_k dk}
{1+2C_{M^*}\bigg\{\int_0^\infty\frac{ n^{\{\kappa\}}_kk^2dk}{E^*(k)}-
\int_0^\infty\frac{n^{\{\kappa\}}_kM^{*2}dk}{E^*(k)}\bigg\}}\;
\end{equation}
where $C_{M^*}=(g_\sigma/m_\sigma)^2\gamma_N/\pi^2\equiv C_S^2/M^2$.
The singularities of $dM^*/dT$ lie in the curve determined by the vanishing of the denominator. Using Eq.~(\ref{ms}), this condition becomes
\begin{equation}\label{dMdT2}
M-2C_{M^*}\int_0^\infty n^{\{\kappa\}}_k\;\frac{dk}{\sqrt{k^2+M^{*2}}}=0\;,
\end{equation}
where
\begin{equation}\label{n1}
n^{\{\kappa\}}_k=\frac{1}{\tilde{e}_\kappa[\beta(\sqrt{k^2+M^{*2}}-\nu)]+1}\;.
\end{equation}
Note that, for $\kappa=0$, we fully recover Eq.~(18) of Ref.~\cite{TSP}. The number of intersections of the solutions obtained from Eq.~(\ref{dMdT2}) and the self-consistency equation given by Eq.~(\ref{ms}) determines how the decoupling happens. We have a first or a second order phase transition, respectively, for two or one intersections. If there is no intersections, the thermodynamical behavior is continuous. The numerical results shown in Fig. \ref{fig8}(a) can be summarized as follows:

\begin{enumerate}
\item For $\kappa$ lying below the $\kappa\times C_S^2$ curve the phase transitions
are of the first order.
\item For $\kappa$ lying on the $\kappa\times C_S^2$ curve the phase transitions
are of the second order. The corresponding values of temperatures and effective
masses are shown in Figs. \ref{fig8}(b) and \ref{fig8}(c).  
\item For $\kappa$ lying above the $\kappa\times C_S^2$ curve the decoupling is continuous.
\end{enumerate}
    Thus, the order of phase transitions depends not only on the actual
values of $C_S^2$ but also on the values of $\kappa$. 

\section{final remarks}\label{fr}

In this paper, we have investigated the effects of the non-gaussian Kaniadakis
framework on the mean field theory of Walecka (QHD-I) \cite{SW}.  We have used, 
instead of the standard Fermi-Dirac nucleon and antinucleon distribution functions, 
the $\kappa$-quantum distribution obtained by Kaniadakis in the framework of the 
{\it kinetic interaction principle} in Ref. \cite{k1}.

We have considered pure neutron and nuclear matter at nonzero and zero baryon densities. 
In the first case, the non-gaussian effects on nuclear and pure neutron matter, for a  
considerable range of temperature, is to make the equation of state stiffer and to increase 
the intensity of the vector and scalar meson fields, with a consequent lowering of the nucleon 
effective mass (for increasing values of the parameter $\kappa$). We believe that 
it may have consequencies in astrophysical studies, mainly in what concerns the 
calculation of masses of compact objects, such as protoneutron stars.

Another interesting feature of the $\kappa$-QHD-I is that, at temperatures in the range $18~{\rm MeV}<T<20~{\rm MeV}$, phase transitions of first order can be avoided by a convenient variation of the parameter $\kappa$, which allows the determination of a critical $\kappa_c$ parameter at the turning point of an isotherm at a given $T$. 

In the second case, we have examined the phase structure of nuclear matter at high temperature and at zero baryon density. The effective Lagrangian of QHD-I theory is considered for the same set of values of the coupling constants in the coupling constant-plane of Ref.~\cite{TSP}. Given that, at vanishing baryon density, the only nonzero coupling constant is that of the scalar field, a relation between the coupling constant $C_S^2$ and the parameter $\kappa$ 
is obtained. This relation determines, in the $\kappa \times C_S^2$ plane, regions of different thermodynamical 
behaviors.

\vspace{0.5cm}

\acknowledgments  RS and JSA are  partially supported by the Conselho Nacional de Desenvolvimento 
Cient\'{\i}fico e Tecnol\'{o}gico (CNPq - Brazil).


\begin{thebibliography}{30}
\bibitem{text1} As is well known, the simplest model of quantum hadrodynamics is the Walecka model (QHD-I), and although providing a consistent theoretical framework, QHD-I presents some limitations.  In this regard, a more complete theoretical treatment (for zero temperature) was discussed in Ref.~\cite{CFH}, whereas the thermo field dynamics in hot nuclear matter was considered in Ref.~\cite{Hen}. 
Here, however, we limit ourselves to the effects of extensive generalized statistics on the QHD-I theory, as originally developed in Ref.~\cite{SW}.
\bibitem{text} At this limit or above it the convergence of the integrals in 
Eqs. (\ref{ms})-(\ref{press}) breaks down, being necessary the use of the Maxwell 
construction. It is worth emphasizing that this picture comes from QHD-I so that it 
can be very different in other theories.
\end{thebibliography}
\end{document}